\documentstyle[12pt,psfig,epsf]{article}

\newcommand{\eq}[1]{(\ref{#1})}
\newcommand{\nn}{\nonumber}
\newcommand{\fr}{\frac}
\def\Journal#1#2#3#4{{#1} {\bf #2} (#4) #3}
\def\PLB{{\em Phys. Lett.}  B}
\def\PRL{\em Phys. Rev. Lett.}
\def\PRD{{\em Phys. Rev.} D}

\def\PTP{\em Prog.~Theor.~Phys.}

\begin{document}
\topmargin 0pt
\oddsidemargin 1mm
%\begin{titlepage}
%\begin{flushright}
% October 1999
%\end{flushright}

\begin{center}
{\Large   Upper and lower bounds of the lightest CP-even Higgs boson\\ 
                    in the two-Higgs-doublet model
          \footnote{Talk given by Shinya Kanemura 
                    (kanemu@particle.physik.uni-karlsruhe.de)
                    at the 2nd ECFA/DESY Linear Collider Workshop 
                    in Obernai, France (16.-19. October 1999) under the  
                    title {\it Mass bounds of the lightest CP-even Higgs 
                               boson in the two-Higgs-doublet model} }
                      }
\end{center}
 
\begin{center}
{\large   Shinya KANEMURA$^\ast$, 
             Takashi KASAI$^\dagger$, 
             and Yasuhiro OKADA$^\dagger$}\\
\vspace*{6mm}
{\em  $^\ast$ Institut f\"{u}r Theoretische Physik, 
            Universit\"{a}t Karlsruhe\\
             D-76128 Karlsruhe, Germany}\\
{\em $^\dagger$ 
       Theory Group, KEK\\
       Tsukuba, Ibaraki 305-0801, Japan}\\
\end{center}
\vspace{1cm}

%\noindent 
%{\large \bf  Introduction}
   By imposing validity of the perturbation and stability of vacuum 
   up to an energy scale $\Lambda$ ($\leq 10^{19}$ GeV), we evaluate 
   mass bounds of the lightest CP-even Higgs-boson mass ($m_h$) in 
   the two-Higgs-doublet model (2HDM) with a softly-broken discrete 
   symmetry \cite{kko}. 
   In the standard model (SM), both the upper and the lower bounds 
   have been analyzed from these kinds of requirement as a function of 
   $\Lambda$ \cite{lindner,sm-2loop}. 
   There have already been several works on the Higgs mass bounds in the 2HDM 
   without the soft-breaking term \cite{kkt,sher2}.
   Our analysis is a generalization of these works to the case 
   with the soft-breaking term. Because the introduction of the soft-breaking 
   scale changes property of the 2HDM, it is very interesting to see  
   what happens for the mass bounds in this case. 
   Our results are qualitatively different from the previous works 
   in the region of the large soft-breaking mass, 
   where only one neutral Higgs boson becomes light. 
   We find that, while the upper bound is almost the same as in the SM, 
   the lower bound is significantly reduced.
   In the decoupling regime where the model behaves like the SM at low energy, 
   the lower bound is given, for example, by about 100 GeV for 
   $\Lambda = 10^{19}$ GeV and $m_t = 175$ GeV, which is smaller by about 
   40 GeV than the corresponding lower bound in the SM. 
   In general case, the $m_h$ is no longer bounded from below by these 
   conditions. 
   If we consider the experimental $b \rightarrow s \gamma$ constraint, 
   small $m_h$ are excluded in Model II of the 2HDM. 

%\noindent
%{\large \bf The 2HDM}

The Higgs potential of the 2HDM is given for both Model I and Model II as 
\cite{kko}  
\begin{eqnarray}
  {V}_{\rm 2HDM}  &=&     m_1^2 \left| \varphi_1 \right|^2 
                          + m_2^2 \left| \varphi_2 \right|^2 - 
                              m_3^2 \left( \varphi_1^{\dagger} \varphi_2 
                                + \varphi_2^{\dagger} \varphi_1 \right) 
                                \nn  
                       + \frac{\lambda_1}{2} 
                               \left| \varphi_1 \right|^4 
                             + \frac{\lambda_2}{2} 
                               \left| \varphi_2 \right|^4 \nn \\
& &                          + \lambda_3 \left| \varphi_1 \right|^2 
                                \left| \varphi_2 \right|^2 
                      + \lambda_4 
                               \left| \varphi_1^{\dagger} \varphi_2 \right|^2
                             + \frac{\lambda_5}{2} 
                             \left\{ 
                               \left( \varphi_1^{\dagger} \varphi_2 \right)^2
                            +  \left( \varphi_2^{\dagger} \varphi_1 \right)^2
                             \right\}.  \label{pot}
\end{eqnarray}
We here take all the self-coupling constants and the mass parameters 
in \eq{pot} to be real. In  Model II, $ \varphi_1 $ has couplings
with down-type quarks and leptons and $ \varphi_2 $ with up-type quarks.  
Only  $ \varphi_2 $ has couplings with fermions in Model I.

The masses of the charged Higgs bosons $(\chi^\pm)$ and CP-odd Higgs boson 
$(\chi_2)$ are expressed as 
$m_{\chi^\pm}^2 = { M^2} - (\lambda_4 + \lambda_5) v^2 / 2$, and 
$m_{\chi_2}^2 =  { M^2}  - \lambda_5 v^2$, respectively, where 
$M = m_3 / \sqrt{\cos \beta \sin \beta}$, 
$\tan \beta = \langle  \varphi_2 \rangle / \langle  \varphi_1 \rangle$ and 
$v = \sqrt{2}
\sqrt{ \langle  \varphi_1 \rangle^2 + \langle  \varphi_2^2 \rangle} \sim 246$ 
GeV.
The two CP-even Higgs boson masses are obtained by diagonalizing the  
$2 \times 2$ matrix, where each component is given by
$  M_{11}^2 = v^2 \left(\lambda_1 \cos^4 \beta + \lambda_2 \sin^4 \beta 
                    + \frac{\lambda}{2} \sin^2 2 \beta \right)$, $ 
  M_{12}^2 = M_{21}^2 = v^2  \sin 2 \beta 
       \left( - \lambda_1 \cos^2 \beta + \lambda_2 \sin^2 \beta 
                    + \lambda \cos 2 \beta \right)/2$ and  $ 
  M_{22}^2 = v^2 \left(\lambda_1  + \lambda_2  
                    - 2 \lambda \right) \sin^2 \beta \cos^2 \beta 
              + M^2$, 
where $ \lambda \equiv  \lambda_3 + \lambda_4+ \lambda_5$.
The mass of the lighter (heavier) CP-even Higgs boson $h$ ($H$) 
is then given by 
$m_{h,H}^2 =  \left\{ M_{11}^2 + M_{22}^2 \mp 
\sqrt{(M_{11}^2 - M_{22}^2)^2 + 4 M_{12}^4 }\right\}/2$.
For the case of $v^2 \ll M^2$, they can be expressed by  
\begin{eqnarray}
\!\!\!\!\!\!\!\!  m_h^2 &=& 
v^2 \left(\lambda_1 \cos^4 \beta + \lambda_2 \sin^4 \beta 
                    + \frac{\lambda}{2} \sin^2 2 \beta \right) 
+ {\cal O}(\frac{v^4}{M^2}), \label{h-mass-2hdm} \\
\!\!\!\!\!\!\!\!  m_H^2 &=&  
               { M^2} + v^2 \left(\lambda_1  + \lambda_2  
                    - 2 \lambda \right) \sin^2 \beta \cos^2 \beta
+ {\cal O}(\frac{v^4}{M^2}). 
\end{eqnarray}

Notice that the soft-breaking parameter $M$ characterizes the model. 
In the case of $M^2 \gg \lambda_i v^2$, these heavy Higgs bosons 
but the lightest decouple from the low-energy observable, 
and below the scale $M$ the effective theory is the SM with one Higgs doublet. 
On the other hand, if $M^2 \sim \lambda_i v^2$, the masses are controlled 
by the self-coupling constants, and thus the heavy Higgs bosons do not 
decouple and the lightest CP-even Higgs boson can have a different property 
from the SM Higgs boson \cite{kanemu}.

%\noindent
%{\large \bf Analyses}

As the condition of validity of perturbation theory, 
we here require that the running coupling constants of the Higgs 
self-couplings and the Yukawa couplings do not blow up below 
a certain energy scale $\Lambda$: this leads the constraints on 
the coupling constants;  
\begin{eqnarray}
  \forall  \, \lambda_i(\mu) < 8 \pi, \; y_t^2(\mu) < 4 \pi \,, 
       \; (\mu < \Lambda).  \label{blowup}
\end{eqnarray}
Next, from the condition of the vacuum stability we obtain 
constraints; 
\begin{eqnarray}
&& \;\;\;\; \lambda_1(\mu) > 0, \;\; \lambda_2(\mu) > 0, \nn \\
&& \sqrt{\lambda_1(\mu) \lambda_2(\mu)} + \lambda_3(\mu) + 
    \rm{min} \left[ 0,\lambda_4(\mu) + \lambda_5(\mu), 
               \lambda_4(\mu) - \lambda_5(\mu) \right]\, >\, 0 \,, \; 
        (\mu < \Lambda).\label{stable}
\end{eqnarray}
We assume that the tree-level Higgs potential at the weak scale does 
not have any global minimum except for the one we consider: 
there is no CP nor charge breaking at the minimum \cite{kko}. 
The conditions \eq{blowup} and \eq{stable} constrain low-energy 
coupling constants through renormalization group equations (RGE's). 
Thus the mass bounds of the Higgs boson are obtained.

In the decoupling regime ($M^2 \gg \lambda_i v^2$),    
the 2HDM effectively becomes the SM with one Higgs doublet below $M$. 
In order to include this effect, we use the one-loop SM RGE below $M$, 
and the one-loop 2HDM RGE~\cite{inoue} above $M$. 
They are connected at $M$ by identifying the lightest CP-even Higgs 
boson in the 2HDM as the SM one in the mass formulas in both the models. 
We here use this procedure for the case $M^2 \sim \lambda_i v^2$ too,  
because the correction from the SM RGE is numerically very small in this case, 
although this procedure is not really justified there.    

The 2HDM receives rather strong experimental constraints 
from the low energy precision data, 
especially on the $\rho$ parameter. The extra contribution 
of the 2HDM to the $\rho$ parameter should satisfy 
$\Delta \rho_{\rm 2HDM} = - 0.0020 - 0.00049 \frac{m_t - 175 {\rm GeV}}
        {5 {\rm GeV} } \pm 0.0027$. 
Another important experimental constraint comes from the 
$b \rightarrow s\gamma$ measurement \cite{bsg-exp}:     
there is a strong constraint on the charged-Higgs boson mass from below 
by this process in Model II, while Model I is not strongly constrained. 
We examine the general mass bounds of $h$ as a function of $\Lambda$ 
varing all the free parameters under these experimental constraints.

%\noindent
%{\large \bf Results}

\begin{figure}[t]
\centering{
\leavevmode
\psfig{file=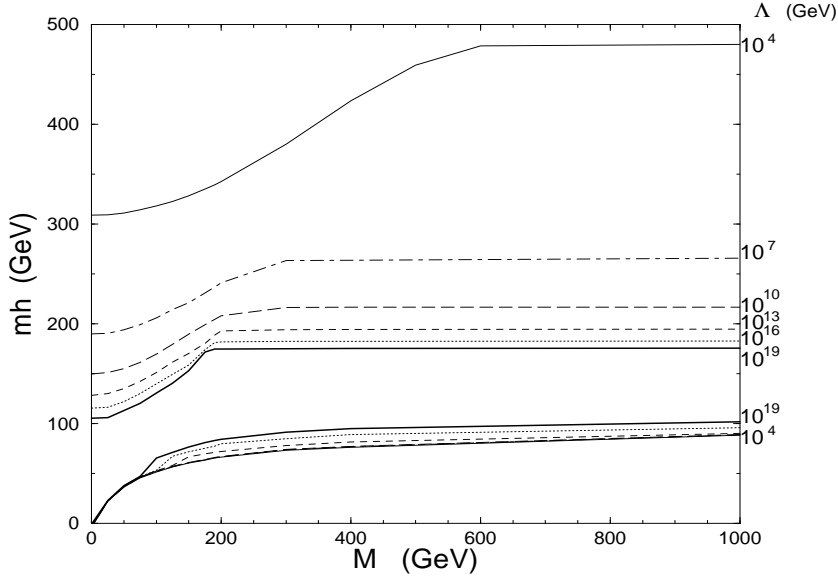,height=80mm,width=110mm,angle=-90}\\
}
\caption{ \small
The mass bounds of the lightest CP even Higgs boson mass 
as a function of $M$ for various $\Lambda$ 
in the Model I 2HDM at $m_t=175$ GeV.}
\end{figure}
\begin{figure}[t]
\centering{
\leavevmode
\psfig{file=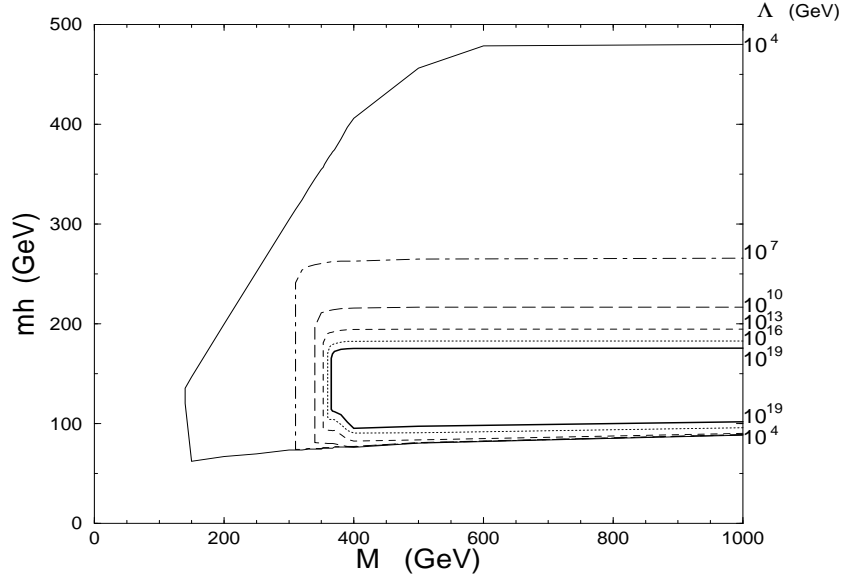,height=80mm,width=110mm,angle=-90}
\vspace{5mm}
}
\caption{\small The mass bounds of the lightest CP even Higgs boson mass 
as a function of $M$ for various $\Lambda$ in the Model II 2HDM 
at $m_t=175$ GeV. Small $M$ region is excluded by 
the $b \rightarrow s \gamma$ results.
\label{fig:t175}
}
\end{figure}

By looking at the RGE's the qualitative result may be understood.  
In decoupling regime, from Eq.~\eq{h-mass-2hdm} we have  
$m_h^2 \sim \lambda_2 v^2$ for $tan \beta \gg 1$.  
The RGE for $\lambda_2$ is given by
\begin{equation}
16 \pi^2 \mu \frac{d \lambda_2}{d \mu}  =  
 12 \lambda_2^2  
 - 3 \lambda_2 (3 g^2 + g'^2) 
 + \fr{3}{2} g^4 + \fr{3}{4} (g^2 + g'^2)^2    
+ 12 \lambda_2 y_t^2 - 12 y_t^4 + A,   \label{lam2-approx} 
\end{equation}
where $A = 2 \lambda_3^2 + 
 2 (\lambda_3 + \lambda_4)^2 + 2 \lambda_5^2 > 0$.
The SM RGE for $\lambda_{SM} (\equiv {m_H^{SM}}^2/v^2)$ takes the same form 
as  Eq.~\eq{lam2-approx}  
substituting $\lambda_{SM}$ and $y_t^{SM}$ to $\lambda_2$ and $y_t$   
and neglecting the $A$ term in the RHS. 
Hence the only difference from the SM RGE is the existence of the 
positive $A$ term, which works to keep the stability of vacuum. 
Thus the lower bound is expected to be reduced in the 2HDM 
in comparison with the SM results.

In Fig.~1 and 2, the upper and lower bounds of the $m_h$ 
are shown as a function of $M$ for various cut-off $\Lambda$  
for Model I and II, respectively.  
In Fig.~1, the allowed region of $m_h$ lies around $m_h \sim M$ 
for $M^2 \ll \lambda_2 v^2$, where the $m_h$ comes from $M_{22} \sim M$ 
and the heavier Higgs boson mass $m_H$ has the mass of 
$M_{11} \sim \sqrt{\lambda_2} v$. At $M = 0$, though there are 
the upper bounds of $m_h$ for each $\Lambda$,  $m_h$ is not bounded 
from below by our condition. Our results at $M = 0$ are consistent 
with Ref.~\cite{sher2}.  
On the other hand, in the decoupling regime ($M^2 \gg \lambda_2 v^2$), 
the situation is reversed; $m_h \sim  M_{11} \sim \sqrt{\lambda_2} v$, 
and the bounds no longer depend on $M$. 
If we take account of the experimental result of $b \rightarrow s \gamma$,
$m_h$ is bounded from below in the Model II as seen in Fig.~2,   
because the small $M$ region necessarily corresponds to small $m_{\chi^\pm}$ 
and this is excluded by the $b \rightarrow s \gamma$ constraint.

\begin{figure}[t]
\centering{
\leavevmode
\psfig{file=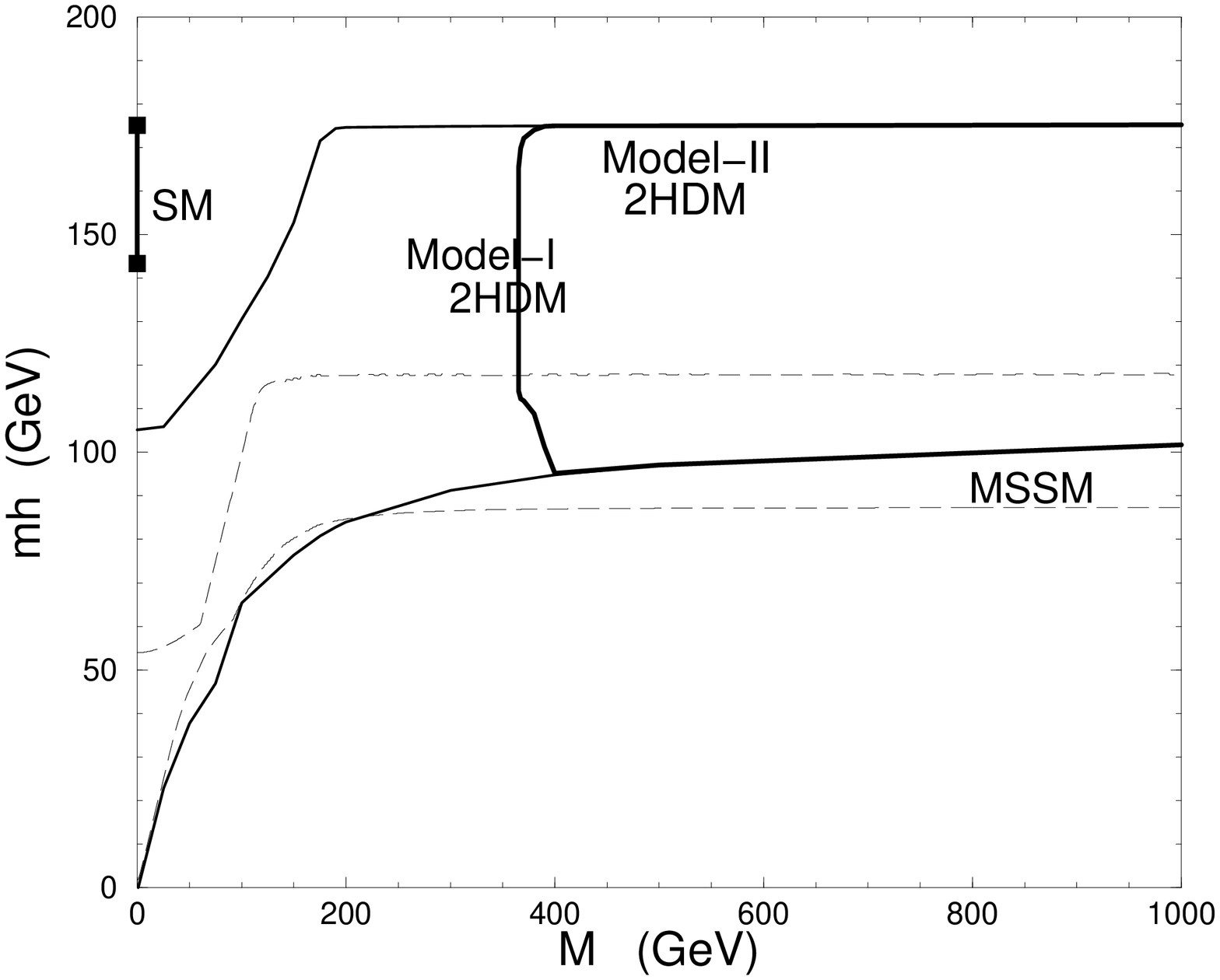,height=80mm,width=110mm,angle=-90}
\vspace{5mm}
}
\caption{\small The mass bounds of the lightest CP even Higgs boson 
in the Model I and II 2HDM as well as of the SM Higgs boson  
for $\Lambda=10^{19}$ GeV. As a reference, the bounds of 
the lightest Higgs boson mass are also shown in the MSSM at the 1 TeV stop 
mass, in which $M$ corresponds to CP-odd Higgs boson mass. 
}
\end{figure}

Finally, we combine the results in the SM and the 2HDM (Model I and II) 
(Fig.~3). We here choose, as an example, $\Lambda = 10^{19}$ GeV 
for comparison of the results in the SM and the 2HDM at $m_t = 175$ GeV.  
For a reference, the bounds of the lightest CP-even Higgs mass 
in the MSSM are also given for the 1 TeV stop mass~\cite{OYYRGE}. 
(In the MSSM, $M$ corresponds to the CP-odd Higgs boson mass exactly.) 
From Fig.~3 it is easy to observe the difference of the bounds 
among the SM, the 2HDM(I) and the 2HDM(II).  
While the upper bounds are all around 175 GeV in these models, 
the lower bounds are completely different as we expect; 
about 145 GeV in the SM, about 100 GeV 
in the Model II (with respect to $b \rightarrow s \gamma$ constraints\footnote{
If we use more conservative way to add theoretical uncertainties for the 
$b \rightarrow s \gamma$ evaluation, the bound on the charged Higgs boson 
or on the $M$ in Model II becomes rather smaller \cite{pol}. 
The lower bound of $m_h$ due to the $b \rightarrow s \gamma$ constraint  
is then reduced for a few GeV according to the changed allowed region of $M$.
} ) and no lower bound in Model I. 
Although we have shown figures in which $m_t = 175$ GeV is taken, the top mass 
dependence cannot be neglected especially for the lower bounds \cite{kko}. 
For example, the lower line for $\Lambda = 10^{19}$ GeV in the 2HDM shown 
in Fig.3 shifts to lower (upper) by 9 GeV for $m_t= 170$ $(180)$ GeV at 
$M = 1000$ GeV. 

In the SM, the next-to-leading order analysis of the effective potential 
shows that the lower bound reduces by about 10 GeV ($\Lambda = 10^{19}$ GeV) 
\cite{sm-2loop}. It may be then expected that a similar reduction of the 
lower bound would occur in the 2HDM by such higher order analysis.

%\noindent
%{\large \bf Discussion}
   In the decoupling regime, the properties of the lightest Higgs boson 
   such as the production cross section and the decay branching ratios 
   are almost the same as the SM Higgs boson. 
   We have not explicitly considered constraint from the 
   Higgs boson search at LEP II~\cite{pol}, but if the Higgs boson is 
   discovered with the mass around $100$ GeV at LEP II or Tevatron 
   experiment in near future and its property is quite similar to the
   SM Higgs boson, the 2HDM with very high cut-off scale is another
   candidate of models which predict such light Higgs boson along with 
   the MSSM \cite{OYYRGE,oyy} and its extensions.

   In summary, we have discussed the mass bounds of the $h$ as a function 
   of a cutoff $\Lambda$ by the requirement of perturbativity and 
   vacuum stability in the non-SUSY 2HDM with the softly-broken discrete 
   symmetry. The upper bounds are almost the same as the SM results, 
   while the lower bounds are significantly reduced even for the 
   decoupling regime. In general case, the mass is no longer bounded from 
   below. If we consider the experimental $b \rightarrow s \gamma$ 
   constraint, the very light $h$ is excluded.

\end{document}